\documentclass[11pt,aps,prd,preprint]{revtex4}

\usepackage[dvips]{color}
\usepackage{epsfig}
\usepackage{amsmath}
\usepackage{graphicx}
\usepackage{subfigure}
\begin{document}
\title{ The $QCD$ parameters are affected by power-law Maxwell field}
\author{B. Khanpour}
\email{b.khanpour@stu.umz.ac.ir}
\author{J. Sadeghi}
\email{pouriya@ipm.ir}
\affiliation{Department of Physics, Faculty of Basic Sciences, University of Mazandaran,\\ P. O. Box 47416-95447, Babolsar, IRAN}
\begin{abstract}
 In this paper, we introduce Einstein power-law Maxwell-Scalar gravity with Liouvill potential in n + 1
dimension. The corresponding metric background or solution of above mentioned action lead us to achieve
some thermodynamics quantities of black hole. On the other hand, we take information from $AdS/CFT$
correspondence and investigate some parameters in $QCD$ as jet quenching and entropic force. We note
that two obtained parameters give some important information about $Q\bar{Q}$ moving in $QGP$ media and
dissociation length of quark-antiquark. Finally, we have several figures for the jet quenching parameter and
entropic force. The corresponding figures show how two parameters are affected by power law Maxwell field
with different of $p$.
  \\\\
{\bf Keywords}: $AdS/CFT$ correspondence;  Power-law Maxwell field solution; Jet-quenching parameter; Entropic force.\\
\end{abstract}
\maketitle
\section{Introduction}
Generally, we know that the duality for the first time is introduced by Maldacena\cite{Maldacena:1997re}. According to  the  Maldacena conjecture\cite{Maldacena:1997re},  the strongly coupled gauge theories with conformal symmetries $(CFT)$ in $n$ dimensional are corresponding to theories of gravity in $n+1$ dimensional in anti-de Sitter $(AdS)$ space-time which is known as $AdS/CFT$ correspondence in zero temperature. But, in finite temperature the strongly coupled gauge theories correspond to gravitational theory in $AdS$ black hole \cite{E. Papantonopoulos:2011,G. Horowitz:2012,CasalderreySolana:2011us,Hartnoll:2009sz,McGreevy:2009xe,Iqbal:2011ae,Natsuume:2014sfa}. So, in that case the gravitational theory of  black hole is a thermal system and also it  has  some  thermal properties \cite{Hawking:1974sw}. So, black hole play important role to understand $AdS/CFT$ correspondence at finite temperature.
 In this paper,  we use the $AdS/CFT$ correspondence in power-law Maxwell field system and calculate jet quenching parameter and entropic force. So, for this reason first we are going to explain generally  power-law Maxwell field solution.
As we know the present epoch, the universe has a positive acceleration which is described by the standard Friedmann model \cite{Riess:1998cb,Perlmutter:1998np,Tonry:2003zg,Lee:2001yp,Netterfield:2001yq,Halverson:2001yy,Spergel:2003cb}.  We note that the Einstein's gravity with  dilaton scalar fields and some  low-energy limit of string theory play important role in  several aspect of cosmology. On the other words from low-energy limit of string theory, one can reach to Einstein's gravity along with a dilaton scalar field \cite{M. B. Green,B. Zwiebach,J. Polchinski,K. Becker}. The most important theory in such direction will be  dilaton gravity including power-law Maxwell field term \cite{Zangeneh:2015wia}.

In the other hand, because of breaking conformal symmetry in $QCD$, it seems that such symmetry should be violated in gravity side. For this reason, some authors introduced several ways. For example hard wall and soft wall models \cite{Polchinski:2000uf,Polchinski:2001tt,Karch:2006pv}. Now, if we look at to the Einstein-Dilaton-power law Maxwell field theory, we will see  in the case of scalar field for  particular power of the massless Klein-Gordon Lagrangian  in arbitrary dimensions \cite{Hassaine:2007sv} and also for  electrodynamic Lagrangian in higher dimensions we have conformal invariance. Here we mention that  Lagrangian $(F_{\mu\nu}F^{\mu\nu})^{\frac{n+1}{4}}$ is conformally invariant $(n+1)$-dimensions in Maxwell Lagrangian \cite{Hassaine:2007py}. But, the Maxwell Lagrangian $F_{\mu\nu}F^{\mu\nu}$ is conformally invariant only in four dimensions. This  power-law Maxwell field term in action play important role for the breaking of conformal invariance, so this will be motivation for the employing such theory for the calculating some parameter in $QCD$. As we will see, interesting results are obtained.

Therefore, we take advantage from such action and metric background and discuss two phenomena  in particle physics as a jet quenching and entropic force. First we are going to give some review to the jet quenching and its properties in $QCD$. As we know deconfined quark-gluon plasma ($QGP$'s) are created in ultra relativistic heavy-ion collision BNL, Relativistic Heavy Ion Collider(RHIC) and the CERN Large Hadron Collider(LHC) \cite{Gyulassy:2004zy,Shuryak:2004cy,Muller:2012zq}.
Two of the most striking properties of $QGP$'s are the perfect (minimally viscous) fluidity as quantified by their shear viscosity to entropy density $\frac{\eta}{s}\sim 0.1-0.2$ \cite{Danielewicz:1984ww,Hirano:2005wx,Majumder:2006wi,Song:2008si,Shen:2010uy} and the strong quenching of high energy jets quantified by the normalized jet transport coefficient $\frac{\hat{q}}{T^3}$ \cite{Burke:2013yra,Baier:2000mf}.\\\\\
Also from RHIC and LHC experiments shown that the quark gluon plasma ($QGP$) are  strongly coupled. Thus, one can not use the perturbation theory in such regime . Fortunately, the $AdS/CFT$ correspondence provides a  the suitable method for calculating  the quantum chromodynamics ($QCD$) parameter,  one of these parameters is jet quenching . In particle physics when heavy ions collide to each other, they are fragmented and produced quark-antiquark will end up back-to-back jets. The quarks have to travel a long way into the $QGP$, in that case they will lose energy in this process. This is  jet signal which is  received by the detector, such  phenomenon is called jet quenching. These produced quark-anti quark as a form of jet give us information about the interaction of the fluid and corresponding particle \cite{Liu:2006ug,BitaghsirFadafan:2010zh,Armesto:2006zv,Avramis:2006ip,Lin:2006au,Nakano:2006js,Caceres:2006as,Buchel:2006bv} . Also, second step we discuss on the entropic force in quarkonium system in $QCD$. As we know the concept of entropic forces will be result of many-body phenomena and also will be interesting in several branch of physics. Here we note that the effect of such force arises from thermodynamic  drive of a many-body system to increase its entropy rather than microscopic system. In that case the more evidence for the describing entropic force coming from  Erik Verlinde’s paper \cite{Verlinde:2010hp}.  For the first time he proposed that the gravitational force will be form of entropic force with some isotropic and homogenous background. He extended such theory for the  Einstein equation and also obtained all equation in cosmology. On the other hand, we have gauge theory in section of high energy physics such as $QCD$. This theory included  abelian and non abelian gauge fields which is bridge between all of interaction of nature. Also Erik Verlinde  take  entropic force and extend such topic to abelian, non abelian gauge and different matter fields.

The ref \cite{Satz:2015jsa}, discussed quarkonium binding and entropic force. He studied quarkonium binding in terms of potential and shown that the relevant potential is the free energy different $F(T,r)$ between a medium with and one without  $Q\bar{Q}$ pair. Also he proved that by increasing of the internal energy $U(T,r)$ with increasing separation distance $r$ and completely correspond to repulsive entropic force. So here we understand that the entropic force also play important role in binding energy in $Q\bar{Q}$ \cite{Satz:2015jsa,Fadafan:2015ynz}.
These information give us motivation to take power-law Maxwell field solution and investigate jet quanching parameter and entropic force. Such investigation may be interesting for the description some phenomena in $QCD$ and also cosmology. The structure of the paper as follows;
In section II,  we present general method for the solving the field equation of Einstein - Power law Maxwell-Scalar gravity in $n+1$ dimension. In that case, we consider Liouvill potential and obtain black hole solution. Also, we arrange the parameters of the black hole in the corresponding theory. In section III,  we take Nambu-Goto action with $\tau$, $\sigma$ coordinates
for the parametrization the world-sheet. So, in order to obtain the jet quenching parameter, we employ holographic description which is
 related to the Wilson loop joining two light-like lines. Also, we take the corresponding metric background in  Nambu-Goto action and use  Wilson loop  one can achieve the  jet quenching parameter. In section IV, we study the entropic force quark-antiquark in  distance $L$. In this section, for obtaining the entropic force, one needs to calculate entropy ($S$), temperature ($T$) and quark-antiquark distance
($L$) through $AdS/CFT$ correspondence. Finally in  last section we have some conclusion and suggestion.
\section{A review of  power-law Maxwell field solution}\label{SE}
Now we are going to investigate power-law Maxwell field action. So,
the action for Einstein gravity coupled to a dilaton field with power-law Maxwell filed will be following form,
\begin{eqnarray}
 S&=&-\frac{1}{16\pi}\int{d^{n+1}x\sqrt{-g}\left[\Re-\frac{4}{n-1}(\nabla{\phi})^2-V(\phi)+(-e^{-\frac{4\alpha\phi}{n-1}}F)^p\right]},
\end{eqnarray}
where $\phi$ is dilaton field and $V(\phi)$ is the potential for the dilaton field. $p$ is degree of nonlinearity of Maxwell field and $\alpha$ is strength of coupling of the electromagnetic and scalar field.  In order to have solution for the corresponding action,  one can consider the dilaton potential with
three Liouville-type as \cite{Zangeneh:2015wia},
\begin{eqnarray}
 V(\phi)=2\Lambda_1e^{2\zeta_1\phi}+2\Lambda_2e^{2\zeta_2\phi}+2\Lambda e^{2\zeta_3\phi},
\end{eqnarray}
so  black hole form solution will be following \cite{Zangeneh:2015wia},
 \begin{equation}\label{metric}
ds^{2}=-f(r)dt^{2}+\frac{1}{f(r)}dr^{2}+r^{2}R(r)^2h_{ij}dx^idx^j,
\end{equation}
where
 \begin{eqnarray}\label{f}
 \nonumber
 f(r)&\!\!=\!\!&
 \frac{k(n-2)(1+\alpha^2)^2 r^{2\gamma}}{(1-\alpha^2)(\alpha^2+n-2)b^{2\gamma}}-\frac{m}{r^{(n-1)(1-\gamma)-1}}-\frac{2\Lambda b^{2\gamma}(1+\alpha^2)^2 r^{2(1-\gamma)}}{(n-1)(n-\alpha^2)}\nonumber\\
 &&+\frac{2^pp(1+\alpha^2)^2(2p-1)^2 b^{-\frac{2(n-2)p\gamma}{2p-1}}q^{2p}}{\Pi(n+\alpha^2-2p)r^{\frac{2[(n-3)p+1]-2p(n-2)\gamma}{2p-1}}},\nonumber\\
\end{eqnarray}
where $k$ can be -1, 0, 1 and $b$ is a positive constant and also we have following values for  $\gamma$, $\Pi$, $\zeta_1$,  $\zeta_2$, $\zeta_3$, $\Lambda_1$ and $\Lambda_2$ \cite{Zangeneh:2015wia}
\begin{eqnarray}
 \nonumber
 \gamma&=&\frac{\alpha^2}{\alpha^2+1},~~~~~ \Pi=\alpha^2+(n-1-\alpha^2)p, ~~~~~\zeta_1=\frac{2}{(n-1)\alpha},\nonumber\\
 &&\zeta_2=\frac{2p(n-1+\alpha^2)}{(n-1)(2p-1)\alpha}, ~~~~~\zeta_3=\frac{2\alpha}{n-1},\nonumber\\
 &&\Lambda_1=\frac{k(n-1)(n-2)\alpha^2}{2b^2(\alpha^2-1)}, ~~\Lambda_2=\frac{2^{p-1}(2p-1)(p-1)\alpha^2 q^{2p}}{\Pi b^{\frac{2(n-1)p}{2p-1}}}.
\end{eqnarray}
By using the Einstein equation and field equation, one can arrange $R(r)$  and $\phi(r)$ as \cite{Zangeneh:2015wia},
\begin{eqnarray}
 R(r)=e^{\frac{2\alpha \phi(r)}{n-1}},
\end{eqnarray}
and
\begin{eqnarray}
 \phi(r)=\frac{(n-1)\alpha}{2(\alpha^2+1)}\ln(\frac{b}{r}).
\end{eqnarray}
As we know the holography as a gauge/gravity duality is a powerful to study the $QCD$ and hadron physics. On the other hand the dynamics of a moving quark and the motion of a quark-antiquark pair in a strongly coupled plasma in the context of gauge/gravit also be important  for the particle physics phenomena.
Also, in dual theory, the black hole object play important role for the obtaining some parameter in $QCD$. Because black hole is thermal object and can be source of some temperature and heat. So, we are going to obtain thermodynamics properties, such as Hawking temperature.

First of all we use the  equation  $f(r_+)=0$, and obtain $m$,  which is given by \cite{Zangeneh:2015wia,Mo:2016jqd},
\begin{eqnarray}
 \nonumber
 m&\!\!=\!\!&
 \frac{k(n-2)b^{-2\gamma} r_+^{\frac{\alpha^2+n-2}{\alpha^2+1}}}{(2\gamma-1)(\gamma-1)(\alpha^2+n-2)}-\frac{2\Lambda b^{2\gamma} r_+^{\frac{n-\alpha^2}{\alpha^2+1}}}{(n-1)(\gamma-1)^2(n-\alpha^2)}\nonumber\\
 &&+\frac{2^pp(2p-1)^2 b^{-\frac{2(n-2)p\gamma}{2p-1}}q^{2p}r_+^{-\frac{\alpha^2-2p+n}{(2p-1)(\alpha^2+1)}}}{\Pi(\gamma-1)^2(n+\alpha^2-2p)}.\nonumber\\
\end{eqnarray}
 In order to determine Hawking temperature, we must calculate the $T=\frac{f'(r_+)}{4\pi}$, so one can obtain following equation,
 \begin{eqnarray}
 T=\frac{\alpha^2+1}{4\pi}(A_1 r_+^{2\gamma-1}-A_2 r_+^{1-2\gamma}-A_3 r_+^{-\eta}),
\end{eqnarray}
where
\begin{eqnarray}
 \nonumber
 A_1&=&\frac{k(n-2)}{b^{2\gamma}(1-\alpha^2)},~~~~~ A_2=\frac{2\Lambda b^{2\gamma}}{n-1}, ~~~~~A_3=\frac{2^pp(2p-1) b^{-\frac{2(n-2)p\gamma}{2p-1}}}{\Pi q^{-2p}},\nonumber\\
 &&\eta=\frac{2p(n-2)(1-\gamma)+1)}{2p-1}, ~~~~~\Lambda=\frac{-(n-1)n}{2l^2},
\end{eqnarray}
where   here $\Lambda$ is  the cosmological constant \cite{Zangeneh:2015wia} and we assume the $l=1$. In the holography point of view the cosmological constant play as a pressure.

 The importance of jet quenching and entropic force in $QCD$ lead us to obtain such parameters with use of gauge/gravity duality tools. So we take above information and black holes with power-law Maxwell field and obtain the jet quenching parameter and entropic force.

\section{The effect of power-law Maxwell field on jet quenching parameter in $QCD$}
As we know the  $AdS/CFT$ correspondence help us to obtain some important parameters in $QCD$. For example in such context we calculate the jet quenching parameter, with use of the some metric background. So, in this section we analyze the behavior of the jet quenching
parameter for the power-law metric background (3). Therefore, in order to calculate  the jet quenching parameter, we  generally consider the light-cone metric. We will take the Nambu-Goto action with  $\tau, \sigma$ coordinates for the parametrization the world-sheet. So,  in the holographic
description, the jet quenching parameter is related to the Wilson loop joining two light-like lines with following equation,
\begin{eqnarray}
<\mathcal{W}^A(\textit{C})>=exp(-\frac{1}{4\sqrt{2}}\hat{q}L^-L^2).
\end{eqnarray}

where $\mathcal{W}^A(\mathcal{C})$ is adjoint Wilson loop and $C$ is a null-like rectangular Wilson loop formed a dipole with heavy $Q\bar{Q}$ pair.  The quark and antiquark are
separated by a small length $L$ and travel along the $L^-$ direction. Also, by using relations
\begin{eqnarray}
<\mathcal{W}^F(\mathcal{C})>^2\simeq<\mathcal{W}^A(\mathcal{C})>,
\end{eqnarray}
and
\begin{eqnarray}
<\mathcal{W}^F(\mathcal{C})>=e^{-S_I}
\end{eqnarray}
one can obtained jet quenching parameter with following formula,
\begin{eqnarray}\label{eq15}
\hat{q}=8\sqrt{2}\frac{S_I}{L^-L^2},
\end{eqnarray}
where $<\mathcal{W}^F(\mathcal{C})>$ is Wilson loop in the fundamental representation and $S_I=S-S_0.$ Here $S$ is the total energy of the quark and anti-quark pair and $S_0$ is the self-energy of the isolated quark and anti-quark. In that case  the $S_I$ is the regularized string world-sheet action.

We are going to start  the following metric background  and calculate the  jet quenching parameter,
\begin{eqnarray}
ds^{2}=-f(r)dt^{2}+\frac{1}{f(r)}dr^{2}+r^{2}R(r)^2\Big[h_{11}dx^1dx^1+h_{22}dx^2dx^2+...\Big],
\end{eqnarray}
and apply following light-cone coordinates,
\begin{eqnarray}
x^{\pm}=\frac{t\pm \sqrt{h_{11}}x^1}{\sqrt{2}}.
\end{eqnarray}
so, we have
\begin{eqnarray}
ds^{2}&=&\Big(\frac{r^2 R(r)^2-f(r)}{2}\Big)\Big[(dx^{+})^2+(dx^{-})^2\Big]+\Big(r^2 R(r)^2+f(r)\Big)dx^+dx^- +\frac{dr^2}{f(r)}\nonumber\\
&&+r^2 R(r)^2\Big[h_{22}(dx^2)^2+h_{33}(dx^3)^2+....\Big].
\end{eqnarray}
The static gauge are choose as $\tau=x^{-}$ and $h_{22}x^{2}=\sigma$ by ($0<x^{-}<L^{-}$) and ($\frac{-L}{2}<y<\frac{L}{2}$) limitations.
 We consider quark and anti-quark pair at location of $y=\pm \frac{L}{2}$. In the limit of $L^-\gg L$, one can neglect of $x^-$ dependence of world-sheet. Therefore, in that case the string profile is completely obtained by $r=r(y).$
The only variable on the string world-sheet are $x^{-}=\tau$ and $\sigma=x^2=y$ and the other coordinates as $x^+, x^3, ...$ are constant. Therefore, we have  string induced metric as,
\begin{eqnarray}\label{eq19}
ds^{2}=\Big(\frac{r^2 R(r)^2-f(r)}{2}\Big)(dx^{-})^2+\frac{dr^2}{f(r)}+r^2 R(r)^2 dy^2.
\end{eqnarray}
In the other hand, we take  $r=r(y)$ and  $dr^2=r'^2dy^2,$ where $r'=\frac{dr}{dy}.$ So, one can rewrite $ds^{2}$ as,
\begin{eqnarray}
ds^{2}=\Big(\frac{r^2 R(r)^2-f(r)}{2}\Big)(dx^{-})^2+\Big(r^2 R(r)^2+\frac{r'^2}{f(r)}\Big) dy^2,
\end{eqnarray}
In this case one can arrange the metric as,
\begin{eqnarray}
g_{\alpha\beta}=\left(
  \begin{array}{cc}
    \frac{r^2 R(r)^2-f(r)}{2} & 0 \\
    0 & r^2 R(r)^2+\frac{r'^2}{f(r)} \\
  \end{array}
\right).
\end{eqnarray}
The Nambu-Goto action will be the following,
\begin{eqnarray}
S&=&-\frac{1}{2\pi \alpha'}\int \int_{0}^{L^-} d\tau d\sigma \sqrt{-det g_{\alpha\beta}}\nonumber \\
&&=\frac{L^-}{ \pi \alpha'}\int_{0}^{L/2} dy \sqrt{-det g_{\alpha\beta}},\\\nonumber
\end{eqnarray}
and
\begin{eqnarray}
S=\frac{L^-}{\sqrt{2}\pi \alpha'}\int_{0}^{L/2} dy \sqrt{\Big(f(r)-r^2 R(r)^2\Big)\Big(r^2 R(r)^2+\frac{r'^2}{f(r)}\Big)},
\end{eqnarray}
where $\frac{1}{2\pi\alpha'}$ is string tension and $\alpha'$ is related to the 't Hooft coupling constant with $\frac{1}{\alpha'}=\sqrt{\lambda}$. We consider $AdS$ radius equal one.
In the above formula the lagrangian density does not depend to  $y$ explicitly , then corresponding hamiltonian is conserved and one can write following,
\begin{eqnarray}
\frac{\partial \mathcal{L}}{\partial r'}r'-\mathcal{L}=C,
\end{eqnarray}
where $C$ is the constant energy of motion which is given by,
\begin{eqnarray}
\mathcal{L}=\sqrt{\Big(f(r)-r^2 R(r)^2\Big)\Big(r^2 R(r)^2+\frac{r'^2}{f(r)}\Big)}.
\end{eqnarray}
One can obtained the equation of motion for $r$ as,
\begin{eqnarray}\label{eq26}
r'=rR(r)\sqrt{f(r)\Big[\frac{r^2 R(r)^2\Big(f(r)-r^2 R(r)^2\Big)}{C^2}-1\Big]}.
\end{eqnarray}
Now, we insert this relation into   Nambu-Goto action and we have,
\begin{eqnarray}\label{eq27}
S=\frac{L^-}{\sqrt{2}\pi \alpha'}\int dr \sqrt{\frac{f(r)-r^2R(r)^2}{f(r)}}\Big[1-\frac{C^2}{r^2R(r)^2(f(r)-r^2R(r)^2)}\Big]^{-1/2}
\end{eqnarray}
Two ends of string placed in location of $y=-L/2,~~y=+L/2$. The mentioned  symmetry in string, lead us to have boundary condition as $r(\pm\frac{L}{2})=\infty$ and in turning point we have $r'(0)=0$.

In order to obtainthe solution of  $f(r)$ in the equation \eqref{eq26} has two roots,  we apply $r'=0$ and obtain  two solution for the $f(r)$. First solution is $f(r)=0$ and give us turning point at event horizon $r=r_h$. The second solution is  boundary condition in turning point  which is,
\begin{eqnarray}
f=r^2 R(r)^2+\frac{C^2}{r^2 R(r)^2},
\end{eqnarray}
where $r=r_{min}$ specify turning point near the horizon. We consider $C$ very small and have $f\rightarrow r_{min}^2 R(r)^2$. Also, we note that in \eqref{eq26} the factor in square root is negative near the black hole horizon and is positive near the boundary. Of course, we know that $r'^2$ is physical quantity whihch is always positive and for this reason we have $r=r_{min}$ in low limit of the integral where $r_{min}>r_h$. We come back to Nambu-Goto action \eqref{eq27}, so in  small  limit $C$, we have following,
\begin{eqnarray}\label{eq29}
S=\frac{L^-}{\sqrt{2}\pi \alpha'}\int_{r_{min}}^{\infty} dr \sqrt{\frac{f(r)-r^2R(r)^2}{f(r)}}\Big[1+\frac{C^2}{2r^2R(r)^2(f(r)-r^2R(r)^2)}\Big].
\end{eqnarray}
Since action contains self energies of the quark and anti-quark pair, it has divergency.  In order to eliminate the divergence it should be subtracted by the self energy. For this reason, we consider the quark and anti-quark as a straight string that stretched from boundary to the event horizon. In this case, in \eqref{eq19}, we have $d\sigma^2=dy^2=0$
\begin{eqnarray}
S_0=\frac{2L^-}{2\pi \alpha'}\int_{r_{min}}^{\infty} dr \sqrt{g_{--}g_{rr}},
\end{eqnarray}
and therefore
\begin{eqnarray}\label{eq31}
S_0=\frac{L^-}{\sqrt{2}\pi \alpha'}\int_{r_{min}}^{\infty} dr \sqrt{\frac{f(r)-r^2R(r)^2}{f(r)}}.
\end{eqnarray}
For obtaining  the $S_I$ , we subtract two equations \eqref{eq31} and \eqref{eq29},
\begin{eqnarray}
S_{I}=S-S_0=\frac{L^-}{2\sqrt{2}\pi \alpha'}\int_{r_{min}}^{\infty} dr \sqrt{\frac{f(r)-r^2R(r)^2}{f(r)}}\frac{C^2}{r^2R(r)^2\Big(f(r)-r^2R(r)^2\Big)},
\end{eqnarray}
and
\begin{eqnarray}\label{eq33}
S_{I}=\frac{L^-}{2\sqrt{2}\pi \alpha'}\int_{r_{min}}^{\infty} dr \frac{C^2}{r^2R(r)^2\sqrt{f(r)\Big(f(r)-r^2R(r)^2\Big)}}=\frac{L^-C^2}{2\sqrt{2}\pi \alpha'}I,
\end{eqnarray}
where

\begin{eqnarray}
I=\int_{r_{min}}^{\infty} \frac{dr}{r^2R(r)^2\sqrt{f(r)\Big(f(r)-r^2R(r)^2\Big)}}.
\end{eqnarray}
Now, we try to obtain $C$ in terms of separation parameter from quark  anti-quark pair as $L$. For this, we have $y=L/2$, then from equation \eqref{eq26}, one can write
\begin{eqnarray}
\frac{dr}{dy}=rR(r)\sqrt{f(r)\Big[\frac{r^2 R(r)^2\Big(f(r)-r^2 R(r)^2\Big)}{C^2}-1\Big]}.
\end{eqnarray}
Then one can write $\frac{L}{2}$ as
\begin{eqnarray}
\frac{L}{2}=\int_{r_{min}}^{\infty} \frac{dr~C}{rR(r)\sqrt{f(r)\Big[r^2 R(r)^2\Big(f(r)-r^2 R(r)^2\Big)-E^2\Big]}}.
\end{eqnarray}
and
\begin{eqnarray}
\frac{L}{2C}=\int_{r_{min}}^{\infty} \frac{dr}{rR(r)\sqrt{f(r)\Big[r^2 R(r)^2\Big(f(r)-r^2 R(r)^2\Big)\Big]}}\Big[1-\frac{C^2}{f(r)\Big[r^2 R(r)^2\Big(f(r)-r^2 R(r)^2\Big)\Big]}\Big]^{-\frac{1}{2}}.
\end{eqnarray}
In low limit of  $C$
\begin{eqnarray}
\frac{L}{2C}=\int_{r_{min}}^{\infty} \frac{dr}{rR(r)\sqrt{f(r)\Big[r^2 R(r)^2\Big(f(r)-r^2 R(r)^2\Big)\Big]}}\Big[1+\frac{C^2}{2f(r)\Big[r^2 R(r)^2\Big(f(r)-r^2 R(r)^2\Big)\Big]}\Big].
\end{eqnarray}
we ignore  $C^2$ and then obtain following equation,
\begin{eqnarray}
\frac{L}{2C}=\int_{r_{min}}^{\infty} \frac{dr}{r^2R(r)^2\sqrt{f(r)\Big[\Big(f(r)-r^2 R(r)^2\Big)\Big]}}=I.
\end{eqnarray}
By putting this relation into  \eqref{eq33}, we obtain
\begin{eqnarray}
S_I=\frac{L^-L^2}{8\sqrt{2}\pi\alpha'I}.
\end{eqnarray}
Finally, by using the  relation \eqref{eq15}, one can obtain  jet quenching parameter as,
\begin{eqnarray}
\hat{q}=\frac{1}{\pi\alpha'I}.
\end{eqnarray}

\begin{figure}[t!]
\begin{center}
Fig. 1
\subfigure[The behavior of Jet quenching parameter versus T for $\alpha$=0.6 ,k=0(dot), k=1(dash), k=-1(solid) with p=2 b=l=1, q=0.5, n=4]
{\includegraphics[height=5.45cm, width=6cm]{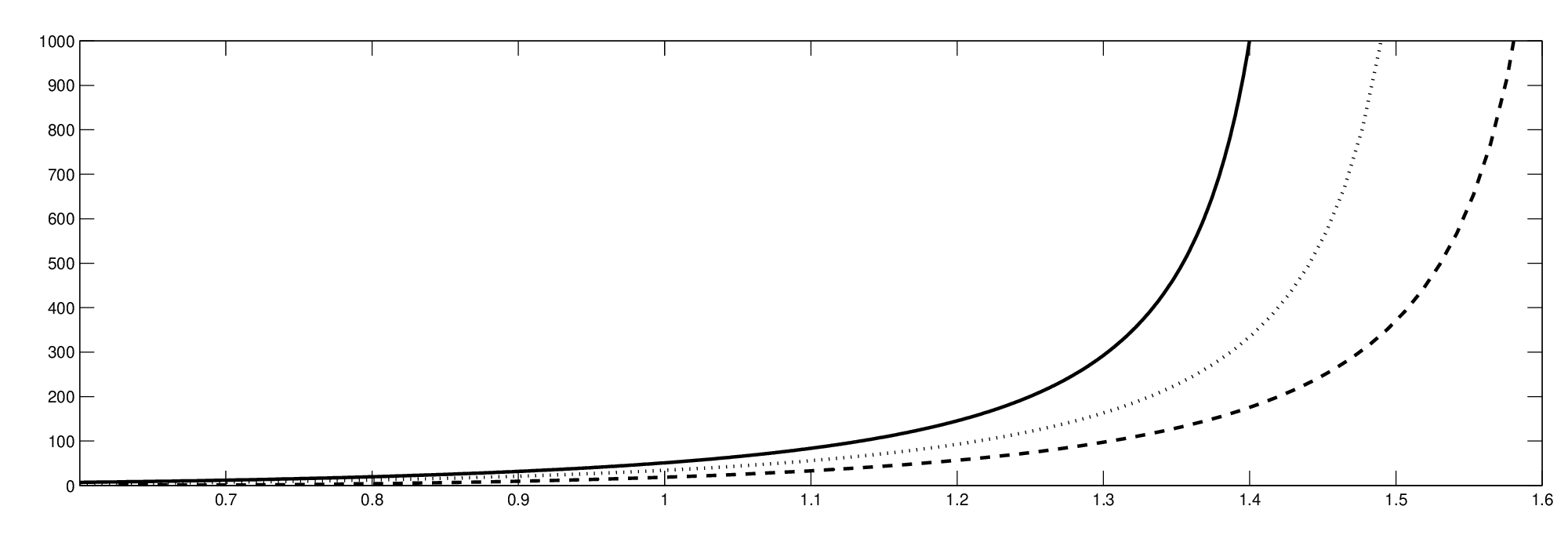}
\label{1}}
\hspace*{1cm}
\subfigure[The behavior of Jet quenching parameter versus $r_+$ for $\alpha$=0.6 ,k=0(dot), k=1(dash), k=-1(solid) with p=2 b=l=1, q=0.5, n=4]
{\includegraphics[height=5cm, width=6cm]{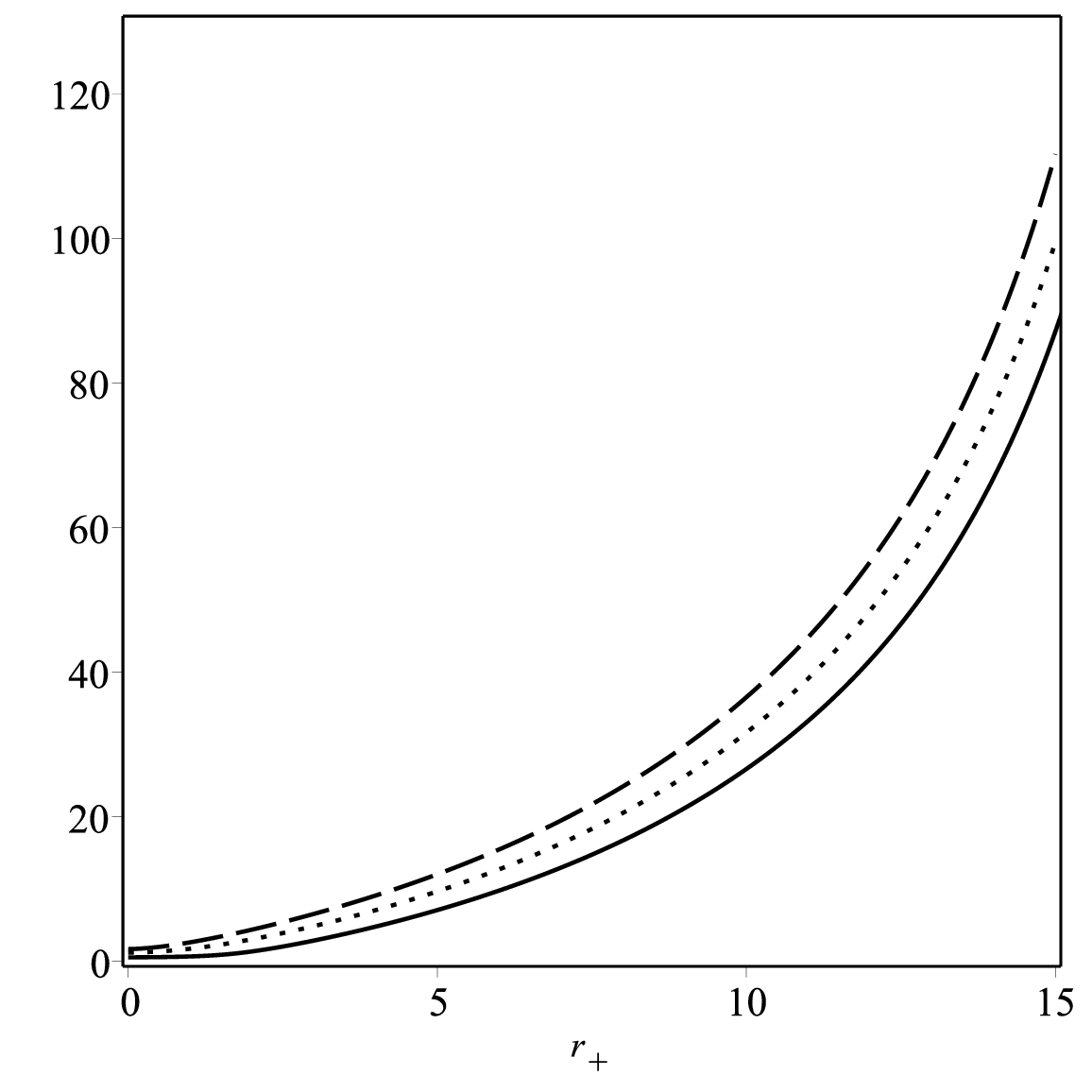}
\label{2}}
\end{center}
\end{figure}

\begin{figure}[t!]
\begin{center}
\subfigure[The behavior of Jet quenching parameter versus T for $\alpha$=0.4 , k=1, q=0.5, n=4, b=l=1 with p=1(solid), p=2(dash)]
{\includegraphics[height=5.45cm, width=6cm]{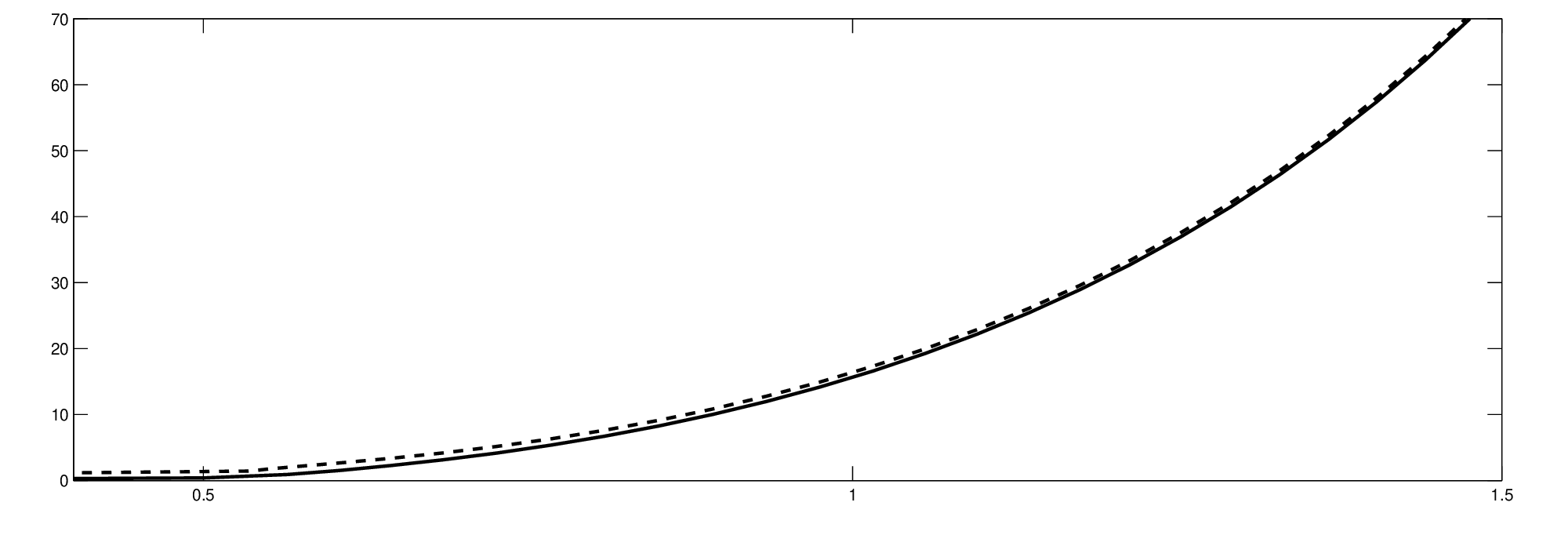}
\label{3}}
\hspace*{1cm}
\subfigure[The behavior of Jet quenching parameter versus T for $\alpha$=0.6 , k=1, q=0.5, n=4, b=l=1 with p=1(solid), p=2(dash)]
{\includegraphics[height=5.4cm, width=6cm]{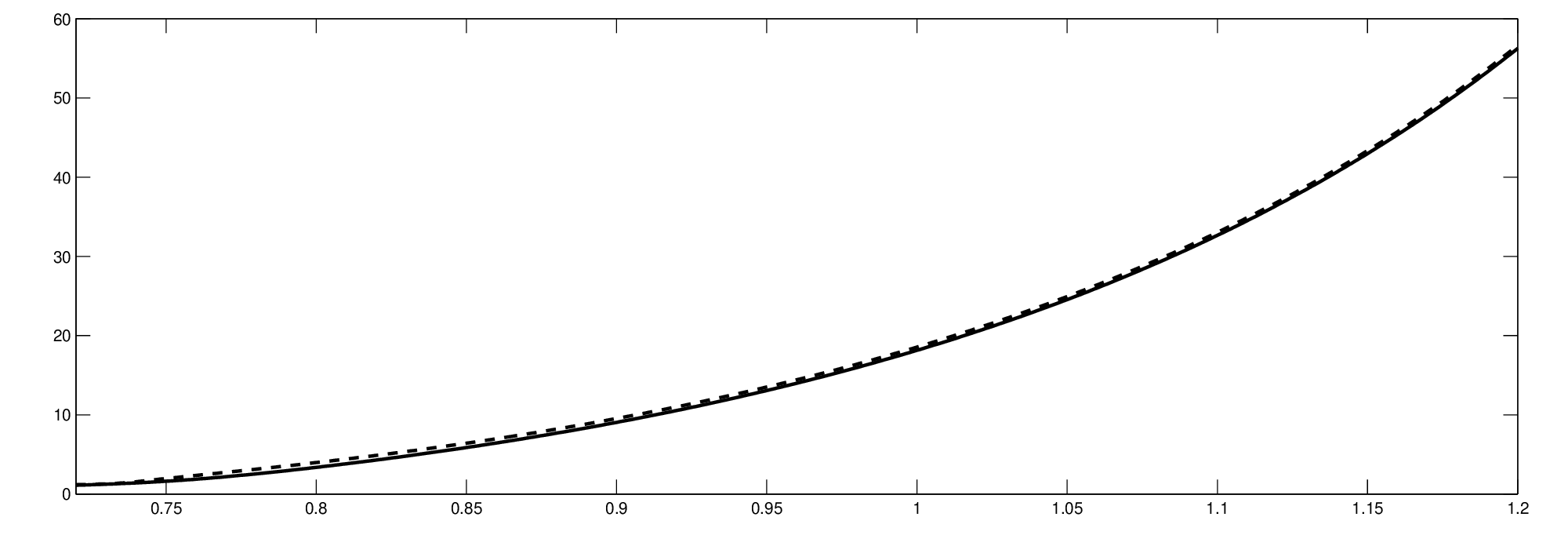}
\label{4}}
\end{center}
\end{figure}
We have some numerically results which are shown in Fig.1 (a,b,c,d). In Fig \ref{1}, Fig \ref{2} we drawn the jet quenching parameters in terms of the black holes temperature and radius of event horizon respectively. These figures  shown that the jet quenching increase with increasing temperature and radius of event horizon. It has also been shown that in fixed temperature, at the positive curvature of the  horizon,  we have  Lowest value jet quenching parameter, and etc.

In Fig \ref{3} and \ref{4} for different $p$ we drawn the jet quenching with respect to temperature. In that case, we choose  $k=1$ and $\alpha=0.4$ and $\alpha=0.6$ respectively. In these figures, we find  important result. The jet quenching parameter is increased by increasing of the power-law Maxwell field $p$. In fact, one can say that the breaking of conformal symmetry, leads to the increasing of jet quenching parameter. By using the table.1, one can compare jet suppression both for conformal and non-conformal.

\begin{table}
\caption{Jet quenching parameter $q_{conformal}$ and $q_{non-conformal}$ in the units of $GeV^2/fm$ for different values of temperature. The temperature is given in the units of $Mev$.}
\vspace{0.3in}
\centering
\begin{tabular}{|c|c|c|c|}
\hline
$T$ & 300 &400 & 500  \\

\hline

$q_{conformal}$&  4.494  &  10.576  &  20.660 \\
\hline

\hline

$q_{non-conformal}$&  4.520  &  10.600  &  20.681 \\
\hline

\end{tabular}
\end{table}

As mentioned before the second parameter was entropic force, this parameter also play important role for the modification of all forces. It also help us to arrange several information in
cosmology. Also, we see such parameter in $QCD$. So for this reason, we will study the effect of the
power-law Maxwell field on the entropic force in following section.

\section{The effect of power-law Maxwell field on entropic force}
In this section, we try to calculate the entropic force for the power-law Maxwell field.
The entropic force is quantity  which is related to the quark-antiquark suppression and is responsible to the separation of quark-antiquark. In fact, increasing of the entropic force leads to the easier separation $Q\bar{Q}$. Our goal is to identify the factors that affect this force. Let us consider a probe that moves in a background, and then we examine the effects of factors such as rapidity and specially power law Maxwell field. Due to power of $p$, conformal symmetry in some cases is breaking. Such breaking symmetry will be appropriate for $AdS/QCD$ correspondence in some place.

The entropic force is obtained by the following formula \cite{Verlinde:2010hp,Kharzeev:2014pha},

\begin{eqnarray}\label{entropy}
\mathcal{F}=T\frac{\partial{S}}{\partial{L}},
\end{eqnarray}
where $L$ is quark-antiquark distance and $T$ is the temperature of the system. Thus, for obtaining the entropic force, one needs to calculate entropy ($S$), temperature ($T$) and quark-antiquark distance ($L$) through AdS/CFT correspondence.

At first, we assume that $QGP$ is at rest and the frame is moving in one direct. In fact, we assume that quark-antiquark is moving with rapidity of $\eta$. Therefore, we have to consider different alignments with respect to the plasma wind, parallel($\theta=0$), transverse($\theta=\pi/2$) and arbitrary direction of the plasma wind. We consider only two first cases. For this reason  we boost the frame in the $x_3$ direction, so that $dt=dt'cosh\eta-dx'_3 sinh\eta$ and $h_{33}dx_3=-dt'sinh\eta+dx'_3cosh\eta.$  Inserting these relations in \eqref{metric} and dropping the primes, we can write:
\begin{eqnarray}
ds^2&=&(-f(r)cosh^2\eta+r^2R(r)^2sinh^2\eta)dt^2+(-f(r)sinh^2\eta+r^2R(r)^2cosh^2\eta)dx_3 ^2 \nonumber\\
&&+2sinh\eta cosh\eta (f(r)-r^2R(r)^2)dt dx_3+\frac{1}{f(r)}dr^2+r^2R(r)^2h_{11}dx_1 ^2+....
\end{eqnarray}

Now,  we  first study ($\theta=\pi/2$) and  choose the static gauge which is  $t=\tau$ and $\sqrt{h_{11}}x_1=\sigma=y$. In this case, quark and antiquqrk are located at $x_1=+L/2$ and $x_1=-L/2$ , so the the  induced metric is given by $dx_2=dx_3=...=0$:
\begin{eqnarray}
ds^2&=&(-f(r)cosh^2\eta+r^2R(r)^2sinh^2\eta)d\tau^2+r^2R(r)^2 d\sigma^2+\frac{1}{f(r)}dr^2,
\end{eqnarray}
where $r=r(\sigma)$, therefore $dr^2=r'^2d\sigma ^2$ and finally the induced metric will be rewritten by following form,
\begin{eqnarray}
ds^2=(-f(r)cosh^2\eta+r^2R(r)^2sinh^2\eta)d\tau^2+(\frac{r'^2}{f(r)}+r^2R(r)^2)d\sigma^2
\end{eqnarray}
We are going to write the Nambu-Goto action of the $U$-shaped string which connected $Q\bar{Q}$ with together in holographic dimension.
\begin{eqnarray}
S=-\frac{1}{2\pi\alpha'}\int d\tau d\sigma \sqrt{-g}.
\end{eqnarray}
It  means that,  one can write,
\begin{eqnarray}\label{action}
S=-\frac{1}{2\pi\alpha'}\int d\tau d\sigma \sqrt{g_1(r)+g_2(r)r'^2},\label{string action}
\end{eqnarray}
where $g_1$ and $g_2$ are equals:
\begin{eqnarray}
&g_1(r)=r^2 R(r)^2(f(r)cosh^2 \eta-r^2 R(r)^2 sinh^2\eta) \nonumber\\
&g_2(r)=cosh^2 \eta+\frac{r^2R(r)^2}{f(r)}sinh^2\eta.
\end{eqnarray}
Action \eqref{action} dose not depend on $\sigma$, thus:
\begin{eqnarray}
\frac{\partial\mathcal{L}}{\partial{r'}}r'-\mathcal{L}=const,
\end{eqnarray}
where $\mathcal{L}=\sqrt{g_1(r)+g_2(r)r'^2}.$

\begin{figure}
\hspace*{1cm}
\begin{center}
\epsfig{file=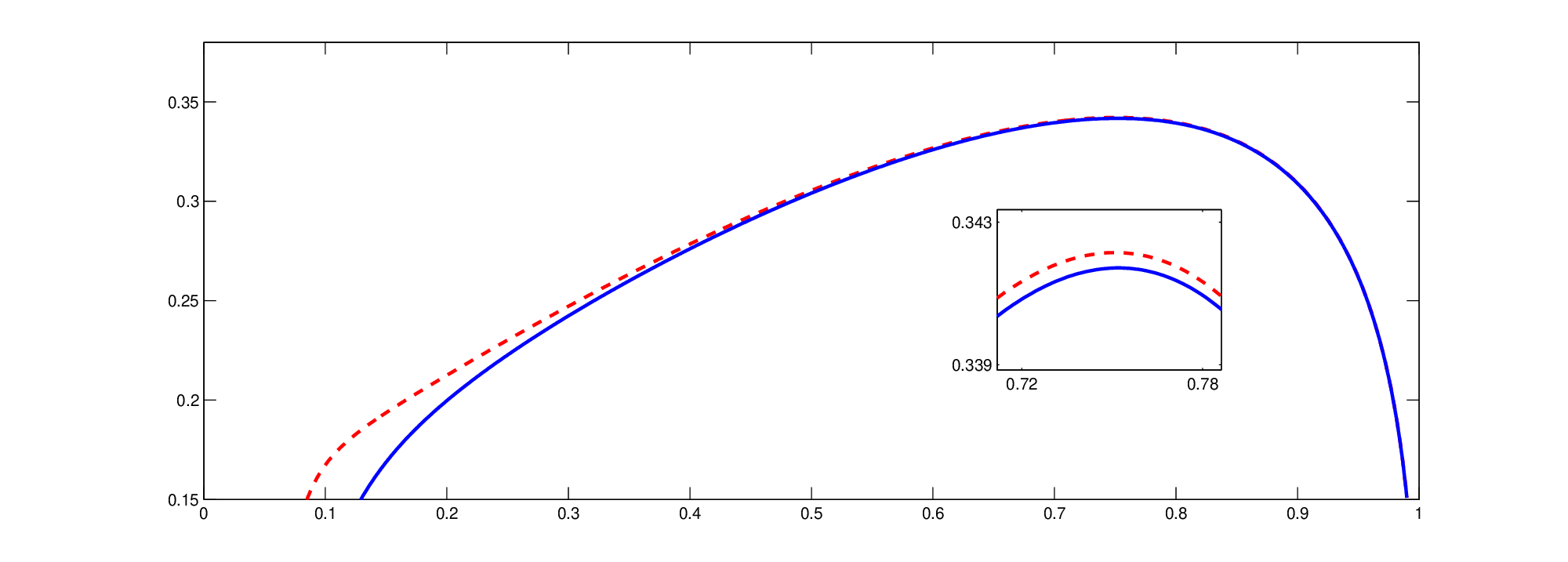,height=6.5cm, width=16cm}
Fig.2, The behavior of LT versus $\varepsilon$ for $\alpha$=0.5 , k=1, q=0.9, $\eta=0$, n=4, b=l=1 with $\theta=\pi/2$ and p=1(dash), p=2(solid)
\end{center}
\end{figure}

By solving above equation, one can obtained $r'$  as
\begin{eqnarray}
\frac{-g_1(r)}{\sqrt{g_1(r)+g_2(r)r'^2}}=const.
\end{eqnarray}
The slope of tangent line for the lowest  point in  $U$ shape of string  is zero, in that case we have  $r=r_c$ and $r'_c=0.$ Therefore,
\begin{eqnarray}\label{r_c}
\frac{-g_1(r)}{\sqrt{g_1(r)+g_2(r)r'^2}}=g_1(r_c)=g_*.
\end{eqnarray}
From \eqref{r_c}, we can obtain $r'$,
\begin{eqnarray}\label{r'}
r'=\sqrt{\frac{g_1(r)^2-g_1(r)g_*}{g_2(r)g_*}},\label{r'}
\end{eqnarray}
where
\begin{eqnarray}
g_*=g_1(r_c)=r_c^2 R(r_c)^2(f(r_c)cosh^2 \eta-r_c^2 R(r_c)^2 sinh^2\eta),
\end{eqnarray}
In this relation the $f(r_c)$ is determined by \eqref{f} with condition of $r=r_c.$  By integrating of \eqref{r'}, the separation  length of the $Q\bar{Q}$ is obtained by,
\begin{eqnarray}
L=2\int_{r_c}^{\infty}{dr}\sqrt{\frac{g_2(r)g_*}{g_1(r)^2-g_1(r)g_*}}.
\end{eqnarray}

\begin{figure}
\hspace*{1cm}
\begin{center}
\epsfig{file=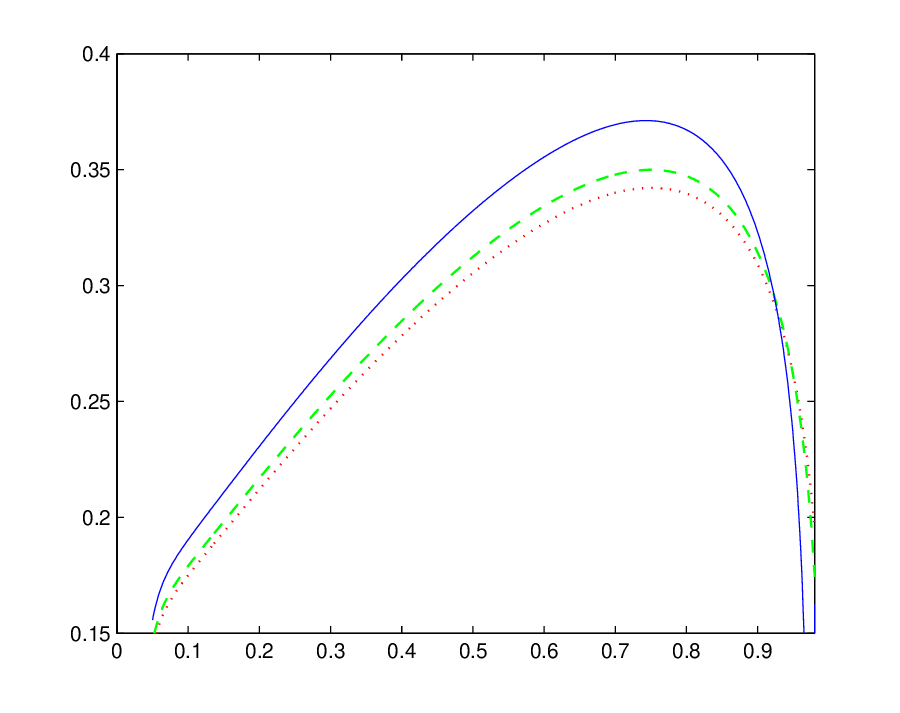,height=6.5cm, width=16cm}
Fig.3, The behavior of LT versus $\varepsilon$ for $\alpha$=0.5 , k=1, q=0.5, n=4, b=l=1, $\theta=\pi/2$  with p=2 and $\eta=0(dot)$, $\eta=0.2(dash)$ and $\eta=0.4(solid)$.
\end{center}
\end{figure}

Now, we are ready draw $LT$ with respect to $\varepsilon=\frac{r_h}{r_c}$, where $r_h<r_c$. In fig(2) we drew $LT$ numerically for different power-law
Maxwell field for fixed rapidity of $\eta$. If we look at deeply to Fig(2),  we will  see that by  increasing $p$  the maximum point of chart  decreases. The maximum  $LT=c$  shows the separation boundary of quark-antiquark. In fact, if $LT>c$ the quarks are screened,  but if $LT<c$ then the fundamental string is connected. So we restrict ourselves to $LT<c$.

Also for fixed $p$ and various rapidity of $\eta$, we have plotted $LT$ with respect to $\varepsilon$. As we can see in Fig(3), the increasing of rapidity lead us to have  large separation boundary. If we contime our calculation for the case of $\theta=0$, we take same results as before.

\begin{figure}
\hspace*{1cm}
\begin{center}
\epsfig{file=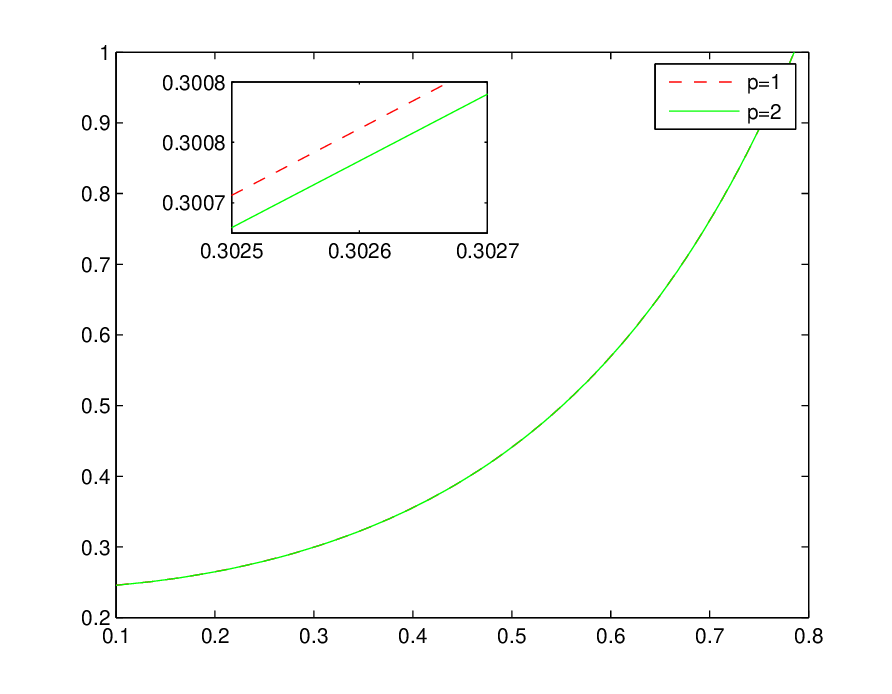,height=6.5cm, width=16cm}
Fig.4, The behavior of $S/\sqrt{\lambda}$ vs $LT$ for $\alpha$=0.5, b=1, q=0.5, k=1, $\eta$=0.9, p=1(dash), p=2(solid)
\end{center}
\end{figure}

\begin{figure}
\hspace*{1cm}
\begin{center}
\epsfig{file=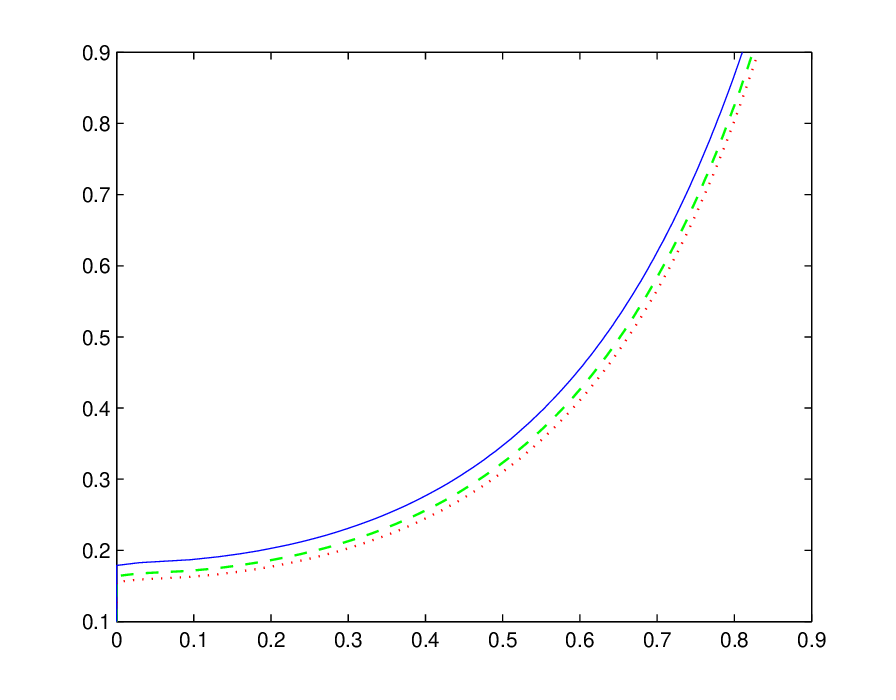,height=6.5cm, width=16cm}
Fig.5, The behavior of $S/\sqrt{\lambda}$ vs $LT$ for $\alpha$=0.5, b=1, q=0.5, k=1, p=2, $\eta$=0(dot), $\eta$=0.3(dash), $\eta$=0.5(solid)
\end{center}
\end{figure}

In the following, we are going to calculate the entropic force. The entropic force can be obtained by the following form,
\begin{eqnarray}
\mathcal{F}=T \frac{\partial{S}}{\partial{L}},
\end{eqnarray}
where $T$ is plasma temperature. In order to calculate the entropic force, we need to obtain equation,

\begin{eqnarray}
S=- \frac{\partial{F}}{\partial{T}},\label{free}\label{entropy2}
\end{eqnarray}
where $F$ is free energy.
As already mentioned, there are two cases. First one, for large values of inter-quarks distance, as ($LT>c$), the quarks are completely disconnected. In that case, if $LT > c$
free energy is not unique \cite{Bak:2007fk}.
One can chooses it as,
\begin{eqnarray}
F=\frac{1}{\pi \alpha'}\int_{r_c}^{\infty}{dr}.
\end{eqnarray}
The corresponding expression leads us the following equation,
\begin{eqnarray}
S=\sqrt{\lambda} \Theta{(L-\frac{c}{T})},
\end{eqnarray}
where $\Theta$ is mathematical function. We leave this case and only consider the second supposition that $LT<c$.

For $\theta=\pi/2$ and $LT<c$ the fundamental string connected. In dual theory for calculating free energy one can utilize on-shell action of fundamental string. So, in that case the equations \eqref{string action} and \eqref{r'} lead us to have following,

\begin{eqnarray}
F=\frac{1}{\pi \alpha'}\int_{r_c}^{\infty}{dr}\sqrt{\frac{g_1(r) g_2(r)}{g_1(r)-g_*}},
\end{eqnarray}
where $g_*=g_1(r_c)$. Using \eqref{entropy2}, numerically one can calculate entropy. By taking $\theta=\pi/2$, fixed rapidity and various of $p$, in fig (4) we plot $\frac{S}{\sqrt{\lambda}}$ with respect to $LT$. As we know, the entropic force, is related to the growth of entropy with distance and is responsible of the dissociating quark-antiquark. By scrutinizing this diagram, we found that slope of the graph in the case of $p=2$ more than of case $p=1$. This means that ,as compared with the conformal case, when the conformal symmetry is broken, entropic force increases. Therefore the breaking of conformal symmetry by the power-law, leads to easier dissociation of $Q\bar{Q}$. In summary, by increasing $p$ the dissociation length will be small.
In figure (5), for $\theta=\pi/2$ and fixed $p$, one can see that increasing of $\eta$ leads to increasing of entropic force. In that case, the dissociation length decreasing. Similar results are obtained by $\theta=0$.

\section{Conclusion}
In this paper, we used black hole solution of Einstein-power law Maxwell-scalar gravity and investigated two important parameter in $QCD$. One of them is jet quenching and the other one is entropic force. These quantities give us important information about moving quark-antiquark in
quark-gluon plasma media and also it’s interaction. The jet quenching shown that how the quark and gluon loses energy in $QGP$. On the other hand, the entropic force related to dissociation length of quark-antiquark in quark-gluon plasma ($QGP$). Our main goal this paper is that, we
want to show the effect of $p$ from power law Maxwell field on the jet quenching and entropic force. In order to see such effects, we drew some graph which shown that the effect of $p$ on the corresponding parameters. The results shown that when the conformal symmetry is broken,
jet quenching parameter and entropic force will increase. It means that due to the conformal symmetry breaking the $QGP$ increase the resistant in front of moving $Q\bar{Q}$ and also reduce the dissociation length. As we know the imaginary part of potential play important role in $QCD$. So, it may be interesting to consider such solution with different $p$ and investigate the imaginary part of potential. Also, one can continue this solution and discuss the relation between the frequency and diffusion constant for the different values of $p$, and find when we have quasi-normal mode.
 \\\\

\newpage

\begin{acknowledgments}
We thank M. Khanpour for helpful discussions about some figures.
\end{acknowledgments}

\end{document}